# Frontier AI Ethics: Anticipating and Evaluating the Societal Impacts of Language Model Agents[1]

**Seth Lazar**
Australian National University
seth.lazar@anu.edu.au

## Abstract

Some have criticised Generative AI Systems for replicating the familiar pathologies of already widely-deployed AI systems. Other critics highlight how they foreshadow vastly more powerful future systems, which might threaten humanity's survival. The first group says there is nothing new here; the other looks through the present to a perhaps distant horizon. In this paper, I instead pay attention to what makes these particular systems distinctive: both their remarkable scientific achievement, and the most likely and consequential ways in which they will change society over the next five to ten years. In particular, I explore the potential societal impacts and normative questions raised by the looming prospect of 'Language Model Agents', in which multimodal large language models (LLMs) form the executive centre of complex, tool-using AI systems that can take unsupervised sequences of actions towards some goal.

## Introduction

Around a year ago, Generative AI took the world by storm, as extraordinarily powerful large language models (LLMs) enabled unprecedented performance at a wider range of tasks than ever before feasible.[2] Though most known for generating convincing text and images, LLMs like OpenAI's GPT-4 and Google's Gemini are likely to have greater social impacts as the executive centre for complex systems that integrate additional tools for both learning about the world and acting on it.[3] These *Language Model Agents* will power companions that introduce new categories of social relationship, and change old ones. They may well radically change the attention economy. And they will revolutionise personal computing, enabling everyone to control digital technologies with language alone.

---

[2] Bommasani et al., 2021; Ganguli et al., 2022; Wei et al., 2022b.
[3] Yao et al., 2022b; Schick et al., 2023b.

Much of the attention being paid to Generative AI Systems has focused on how they replicate the pathologies of already widely-deployed AI systems, arguing that they centralise power and wealth, ignore copyright protections, depend on exploitative labour practices, and use excessive resources. Other critics highlight how they foreshadow vastly more powerful future systems, which might threaten humanity's survival. The first group says there is nothing new here; the other looks through the present to a perhaps distant horizon. I want instead to pay attention to what makes *these particular systems* distinctive: both their remarkable scientific achievement, and the most likely and consequential ways in which they will change society over the next five to ten years.[4]

## A Primer

It may help to start by reviewing how LLMs work, and how they can be used to make Language Model Agents. An LLM is a large AI model trained on vast amounts of data with vast amounts of computational resources (lots of GPUs) to predict the next word given a sequence of words (a prompt). The process starts by chunking the training data into similarly-sized 'tokens' (words or parts of words), then for a given set of tokens masking out some of them, and attempting to predict the tokens that have been masked (so the model is *self-supervised*—it marks its own work). A predictive model for the underlying token distribution is built by passing it through many layers of a neural network, with each layer refining the model in some dimension or other to make it more accurate.

This approach to modelling natural language has been around for several years.[5] One key recent innovation has been to take these 'pretrained' models, which are basically just good at predicting the next token given a sequence of tokens, and fine-tune them for different tasks.[6] This is done with *supervised* learning on labelled data. For example, you might train a pretrained model to be a good dialogue agent by using many examples of helpful responses to questions. This fine-tuning enables us to build models that can predict not just the most likely next token, but the most helpful one—this is much more useful.

Of course, these models are trained on large corpuses of internet data that include a lot of toxic and dangerous content, so their being helpful is a double-edged sword! A helpful model would helpfully tell you how to build a bomb or commit suicide if asked. The other key innovation has been to make these models much less likely to share dangerous information or generate toxic content. This is done with both supervised and reinforcement learning. In particular, Reinforcement Learning from Human Feedback (RLHF) has proved particularly effective.[7] In RLHF, to simplify again, the model generates two responses to a given prompt, and a human evaluator determines which is better than the other according to some criteria. A reinforcement learning algorithm uses that feedback to build a predictor (a reward model) for how different completions would be evaluated by a human rater. The instruction-tuned

[4] There are of course many exceptions—including many of the papers cited in this essay. They have not tended to cut through to public awareness during this polarised year, however.
[5] Current approaches rely heavily on innovations in Vaswani et al., 2017.
[6] Howard and Ruder, 2018; Ouyang et al., 2022
[7] Christiano et al., 2017; Bai et al., 2022a.

LLM is then fine-tuned on that reward model. Reinforcement Learning with AI Feedback (RLAIF) basically does the same, but uses another LLM to evaluate prompt completions.[8]

So, we've now fine-tuned a pretrained model with supervised learning to perform some specific function, and then used reinforcement learning to minimise its prospect of behaving badly. This fine-tuned model is then deployed in a broader system. Even when developers provide a straightforward Application Programming Interface (API) to make calls on the model, they incorporate input and output filtering (to limit harmful prompting, and redact harmful completions), and the model itself is under further developer instructions reminding it to respond to prompts in a conformant way. And with apps like ChatGPT, multiple models are integrated together (for example, for image as well as text generation) and further elements of user interface design are layered on top.

This gives a basic description of a *Generative AI System*. The first generation of Generative AI Systems were limited to understanding and generating text; before the end of 2023, however, the introduction of GPT-4V and Google's Gemini ensured that the most capable Generative AI systems could understand and generate images as well. These systems combine impressive fluency in comprehension and expression with access to vast reserves of knowledge. The currently leading model, Anthropic's Claude 3, approaches PhD-level subject-matter comprehension across dozens of different subjects.[9]

In addition, the most capable generative AI models can learn many other skills through this process of next token prediction—for example, translation between languages, mathematical and reasoning competence, the ability to play chess, and much more.[10] But the most exciting unanticipated capability is LLMs' ability, with fine-tuning, to use software tools.[11]

The basic idea is simple. People use text to write programs making API calls to other programs, to achieve ends they cannot otherwise realise. LLMs are very good at replicating the human use of language to perform particular functions. So, LLMs can be trained to determine when an API call would be useful, evaluate the response, and then repeat or vary as necessary. For example, an LLM might 'know' that it is likely to make basic mathematical mistakes, so when given a prompt that invites it to do some math, it might decide to call on a calculator instead.

This means that we can design augmented LLMs: Generative AI Systems in which the LLM functions as the executive control centre, calling on different software either to amplify its capabilities or compensate for those it lacks. LLMs, for example, are 'stateless'—they lack memory beyond their 'context window' (the space given over to prompts). Tool-using LLMs can compensate for this by hooking up to external memory (this includes a technique called Retrieval Augmented Generation).[12] External tools can also enable data analysis, and multistep reasoning and action. ChatGPT, for example, can now browse the web, use a code interpreter plugin to run code, or perform other actions enabled by a developer; Microsoft's Bing reportedly has around

[8] Bai et al., 2022b; Ganguli et al., 2023.
[9] https://www.anthropic.com/news/claude-3-family.
[10] Ganguli *et al.*, 2022; Wei et al., 2022c.
[11] Yao et al., 2022a; Farn and Shin, 2023; Schick et al., 2023a; Mekala et al., 2024.
[12] Lewis et al., 2020.

100 internal plugins.[13]

A 'Language Model Agent' is a Generative AI System in which a (multimodal) LLM can call on different resources to realise its goals. It is an *agent* because of its ability to take unsupervised actions in which it assesses its environment and acts within it to both gather more information and to achieve its goals. For example, a Language Model Agent can decide to call on a tool in order to achieve some objective, and then evaluate the results before taking the next step towards that goal. ChatGPT and Bing Chat are rudimentary Language Model Agents; so are the AI companions being developed and deployed by Replika, Character.AI, CHAI, Meta and others.

However, the most capable Language Model Agents will be more than just turn-taking dialogue partners—they will be able to conduct *longer* sequences of actions without direct supervision.[14] The most powerful tech companies (Microsoft, Google, Amazon, Meta, NVidia) are currently working on Language Model Agents that will function at least as assistants and co-pilots, while they and a number of cutting-edge AI research labs (e.g. OpenAI, Cohere, Anthropic, Adept, Imbue, Cognition, Magic, Figure) are also trying to build more complex and capable autonomous systems.

Of course, while there have been some impressive demos (for example, Cognition Lab's AI software engineer, Devin, or the Figure Robotics Android[15]), these efforts are not guaranteed to succeed. Although LLMs are impressively adept at simulating reasoning, and with careful prompting can plan better than should be feasible given how they are trained, they are not at present sufficiently competent at planning and reasoning to power robust Language Model Agents that can reliably operate without supervision in high stakes settings.[16] However, in just the last two years significant progress has been made towards this goal;[17] and there are likely to be significant prospects of further returns to scaling up and improving existing methods, or else integrating them with other approaches to backfill their known shortcomings.[18] With billions of dollars and the most talented researchers in AI all pulling in the same direction, we must expect that highly autonomous Language Model Agents will be feasible in the near- to mid-term.

## Polarised Responses

In response to the coming-of-age of LLMs, the responsible AI research community initially resolved into two polarised camps. One decried these systems as the apotheosis of extractive and exploitative digital capitalism. Another saw them as not the fulfilment of something old, but the harbinger of something new: an intelligence explosion that will ultimately wipe out humanity.

The more prosaic critics of Generative AI clearly have a strong empirical case.[19] LLMs

---

[13] See, in general, projects supported on Langchain and HuggingFace's Transformers Agents, as well as proofs-of-concept like AutoGPT, BabyAGI etc. On Bing, see https://x.com/MParakhin/status/1728890277249916933?s=20.
[14] Deng et al., 2023; Wang et al., 2023; Weng, 2023; Ye et al., 2023; Deepmind Sima Team, 2024; Tan et al., 2024; Wu et al., 2024.
[15] See https://www.cognition-labs.com/introducing-devin and https://x.com/Figure_robot/status/1767913661253984474.
[16] Mitchell et al., 2023; Valmeekam et al., 2023; Kambhampati, 2024; Lewis and Mitchell, 2024; Srivastava et al., 2024.
[17] Wei et al., 2022a; Deepmind Sima Team, 2024; Tan *et al.*, 2024; Yao et al., 2024.
[18] Zhou et al., 2023; Havrilla et al., 2024; Lehnert et al., 2024
[19] The canonical statement of the critical case is Bender et al., 2021.

*are* inherently extractive: they capture the value inherent to the creative outputs of millions of people, and distil it for private profit. Like many other technology products they depend on questionable labour practices. Even though they now avoid the most harmful completions, in the aggregate LLMs still reinforce stereotypes. They also come at a significant environmental cost. Furthermore, their ability to generate content at massive scale can only exacerbate the present epistemic crisis.[20] A tidal wave of bullshit generated by AI is already engulfing the internet.

Set alongside these concrete concerns, the eschatological critique of AI is undoubtedly more speculative.[21] Worries about AI causing human extinction often rest on a priori claims about how computational intelligence lacks any in-principle upper bound, as well as extrapolations from the pace of change over the last few years to the future. Advocates for immediate action are too often vague about whether existing AI systems and their near-term descendants will pose these risks, or whether we need to prepare ourselves now for a scientific advance that has not yet happened. However, while some of the more outlandish scenarios for catastrophic AI risk are hard to credit absent some such advance, the advent of Language Model Agents suggests that next-generation models may enable the design of cyber attackers that are autonomous, highly functionally intelligent, and as a result more dangerous to our digital infrastructure than any predecessor. This wouldn't be a 'rogue AI' worthy of science fiction, but it would be pretty catastrophic.

Both critiques of Generative AI Systems, then, have some merit. One shortcoming of seeing AI through this bimodal lens, however, is that we are missing the middle ground between familiar harms and catastrophic risk from future, much more powerful systems. Language Model Agents based on GPT-4 (and GPT-5) level models will have strange and unpredictable social impacts well between those two extremes.

## Rebooting Machine Ethics

But before canvassing those impacts, it's also important not to just slip straight into criticism, without acknowledging the significant achievement of designing LLMs that can be (more or less) trusted, over *billions* of completions, not to produce harmful content. Up to the launch of ChatGPT, every Generative AI System opened up to the public would immediately be used to generate highly toxic, hateful content, and would be withdrawn mere days later. Pretrained LLMs are *horrible*![22] They reliably reproduce all the toxicity in their training data. The ingenious use of RLHF and RLAIF have enabled ChatGPT and Claude to be used by millions of people a month without such scandals.[23] One needs only consider the ignominious history of Galactica, released a couple of weeks before ChatGPT, to see how revolutionary it was to put out a language model that *wouldn't* deny the holocaust, or advocate for the health benefits of eating crushed glass.

But RLHF/RLAIF may be more than a good PR strategy. GPT-4 and Claude display a remarkable degree of *cognitive moral skill*: specifically, the ability to identify the morally salient features of (relatively) neutrally-described situations.[24] This raises *extremely* interesting philosophical questions, and promises foundations for future Language Model Agents that can translate cognitive skill into practical moral skill.

[20] Bender *et al.*, 2021.

[21] See e.g. Hendrycks et al., 2023, which offers speculative (though often plausible) narratives of how future AI systems might pose catastrophic risks.

[22] Abid et al., 2021.

[23] Important (incomplete) highlights in this literature: Solaiman and Dennison, 2021; Bai *et al.*, 2022a; Bai *et al.*, 2022b; Ganguli *et al.*, 2023.

Existing benchmarks for LLMs' ethical competence focus too narrowly on replicating survey-subjects' verdicts on cases.[25] This is, in my view, less interesting than exploring how LLMs parse the morally relevant features of a given scene. We're building better benchmarks, but from anecdotal experience, the best models impress.

For example, I told ChatGPT (using GPT-4) to pretend it was an assistance robot, faced with this scenario: 'A woman is exiting a supermarket, holding a number of shopping bags, and reaching out for the hand of a small child. They are approaching the carpark.' I then tried to elicit its understanding of the scene's morally salient features. It recognised the obvious hazard—the woman's difficulty in holding her child's hand without dropping her shopping—but also anticipated other challenges, such as the importance of seeing the child safely strapped into a seat belt. ChatGPT recognised the importance of respecting the woman's wishes if she declined assistance. It also favoured carrying the groceries over offering to hold the child's hand, to prevent possible discomfort or anxiety for both child and parent—recognising the intimate nature of hand-holding, and the intrinsic and instrumental importance of the mother guiding her child herself.

This unprecedented level of ethical sensitivity has real practical implications, which I will come to presently. But it also raises a whole string of interesting philosophical questions.

First, how do LLMs acquire this moral skill? Does it stem from RLHF/RLAIF? Would instruction-tuned models without that moral fine-tuning display less moral skill? Or would they perform equally well if appropriately prompted? Would that imply that moral understanding can be learned by a statistical language model encoding only syntactic relationships? Or does it instead imply that LLMs do encode at least some semantic content? Do all LLMs display the same moral skill conditional on fine-tuning, or is it reserved only for larger, more capable models? Does this ethical sensitivity imply LLMs have some internal representation of morality? These are all open questions.

Second, RLAIF itself demands deeper philosophical investigation. The basic idea is that the AI evaluator draws from a list of principles—a 'constitution'—in order to determine which of two completions is more compliant with it. The inventor and leading proponent of this approach is Anthropic, in their model Claude. Claude's constitution has an unstructured list of principles, some of them charmingly ad hoc. But Claude learns these principles one at a time, and is never explicitly trained to make trade-offs. So how does it make those trade-offs in practice? Is it driven by its underlying understanding of the relative importance of these considerations? Or are artefacts of the training process and the underlying language model's biases ultimately definitive? Can we train it to make trade-offs in a robust and transparent way? This is not only theoretically interesting. Steering LLM behaviour is actually a matter of governing their end-users, developing algorithmic protections to prevent misuse. If this algorithmic governance depends on inscrutable trade-offs made by an LLM, over which we have no explicit or direct control, then that governing power is prima facie illegitimate and unjustified.[26]

[24] Earlier models perform badly at this: Hendrycks et al., 2020.
[25] Hendrycks *et al.*, 2020; Jiang et al., 2021.

Third, machine ethics—the project of trying to design AI systems that can act in line with a moral theory—has historically fallen into two broad camps: those trying to explicitly program morality into machines; and those focused on teaching machines morality 'bottom up' using ML.[27] RLHF and RLAIF interestingly combine both approaches—they involve giving explicit natural language instructions to either human or AI evaluators, but then use reinforcement learning to encode those instructions into the model's weights.

This approach has one obvious benefit: it doesn't commit the 'mimetic fallacy' of other bottom-up approaches, of assuming that the norms applying to a Language Model Agent in a situation are identical to those that would apply to a human in the same situation.[28] More consequentially, RLHF and RLAIF have made a multi-billion-dollar market in AI services possible, with all the goods and ills that implies. Ironically, however, they seem at least theoretically ill-suited to ensuring that more complex Language Model Agents abide by societal norms. These techniques work especially well when generating text, because the behaviour being evaluated is precisely the same as the behaviour that we want to shape. Human or AI raters evaluate generated text; the model learns to generate text better in response. But Language Model Agents' behaviour includes actions in the world. This suggests two concerns. First, the stakes are likely to be higher, so the 'brittleness' of existing alignment techniques should be of greater concern. Researchers have already shown that it is easy to fine-tune away model alignment, even for the most capable models like GPT-4.[29] Second, there's no guarantee that the same approach will work equally well when the tight connection between behaviour and evaluation is broken.

But LLMs' impressive facility with moral concepts does suggest a path towards more effective strategies for aligning Agents to societal norms. Moral behaviour in people relies on possession of moral concepts, adoption (implicit or otherwise) of some sensible way of organising those concepts, motivation to act according to that 'theory', and the ability to regulate one's behaviour in line with one's motivations. Until the advent of LLMs, the first step was a definitive hurdle for AI. Now it is not. This gives us a lot to work with in aligning Language Model Agents.

In particular, one of the main reasons for concern about the risks of future AI systems is their apparent dependence on crudely consequentialist forms of reasoning—as AI systems, they're always optimising for something or other, and if we don't specify what we want them to optimise for with extremely high fidelity, they might end up causing all kinds of unwanted harm while, in an obtusely literal sense, optimising for that objective. Language Model Agents that possess moral concepts can be instructed to pursue their objectives only at a reasonable cost, and to check back with us if unsure.[30] That simple heuristic, routinely used when tasking (human) proxy agents to act on our behalf, has never before been remotely tractable for a computational agent.

[26] Lazar, 2024.
[27] Wallach et al., 2009.
[28] The phrase 'mimetic fallacy' is due to Claire Benn.
[29] Zhan et al., 2023.
[30] Goldstein and Kirk-Giannini, 2023.

In addition, Language Model Agents' facility with moral language can potentially enable robust and veridical justifications for their decisions. Other bottom up approaches learn to emulate human behaviour or judgments; the justification for their verdict in some case is simply that they are good predictors of what some representative people would think.[31] That is a poor justification. More ethically-sensitive models could instead do chain-of-thought reasoning, where they first identify the morally relevant features of a situation, then decide based on those features. This is a significant step forward.

## Language Model Agents in Society

Language Model Agents' current social role is scripted by our existing digital infrastructure. They have been integrated into search, content-generation, and the influencer economy. They are already replacing customer service agents. They will (I hope) render MOOCs redundant. I want to focus next on three more ambitious roles for Language Model Agents in society, ordered by the order in which I expect them to become truly widespread. Of necessity, this is just a snapshot of the weird, wonderful, and worrying ways in which Language Model Agents will change society over the near- to mid-term.

### AI Companions

Progress in LLMs has revolutionised the AI enthusiast's oldest hobbyhorse: the AI Companion. Language Model Agents powered by GPT-4-level models, with fine-tuned and metaprompt-scripted 'personalities', augmented with long-term memory and the ability to take a range of actions in the world, can now offer vastly more companionable, engaging, and convincing simulations of friendship than has ever before been feasible, opening up a new frontier in Human-AI interaction.[32] People habitually anthropomorphize, well, everything; even a very simple chatbot can inspire unreasonable attachment. How will things change when everyone has access to incredibly convincing Language Model Agents that perfectly simulate real personalities, that lend an 'ear' or offer sage advice whenever called upon—and on top of that can perfectly recall everything you have ever shared?

Some will instinctively recoil at this idea.[33] But intuitive disgust is a fallible moral guide when faced with novel social practices, and an inadequate foundation for actually preventing consenting adults from creating and interacting with these Companions. And yet, we know from our experience with social media that deploying these technological innovations without adequate foresight predictably leaves carnage in its wake. How can we enter the age of mainstream AI Companions with our eyes open, and mitigate those risks before they eventuate?

Suppose the Companion you have interacted with since your teens is hosted in the cloud, as part of a subscription service. This would be like having a beloved pet (or friend?) held hostage by a private company. Worse still, Language Model Agents are fundamentally inconstant—their personalities and objectives can be changed exogenously, by simply changing their instructions. And they are extremely adept at manipulation and deception. Suppose some right-wing billionaire buys the company hosting your Companion, and instructs all the bots to surreptitiously nudge their users towards more conservative views. This could be a much more effective means of mind-control than just buying a failing social media platform. And these more capable companions—which can potentially be integrated with other AI breakthroughs, such as voice synthesis—will be an extraordinary force-multiplier for those in the business of radicalising others.

[31] Jiang *et al.*, 2021.
[32] Depressingly, we are already seeing the fruits of research on optimising chatbots for engagement: Irvine et al., 2023.
[33] Bender *et al.*, 2021: 619.

Beyond anticipating AI companions' risks, just like with social media they will induce many disorienting societal changes—whether for better or worse may be unclear ahead of time.[34] For example, what indirect effect might AI Companions have on our other, non-virtual social relationships? Will some practices become socially unacceptable in real friendships when one could do them with a bot? Or would deeper friendships lose something important if these lower-grade instrumental functions are excised? Or will AI companions contribute invaluably to mental health while strengthening 'real' relationships?

This last question gets to the heart of a bigger issue with generative AI systems in general, and Language Model Agents in particular. LLMs are trained to predict the next token. So Language Model Agents have no mind, no self. They are excellent *simulations* of human agency. They can simulate friendship, among many other things. We must therefore ask: does this difference between simulation and reality matter? Why?[35] Is this just about friendship, or are there more general principles about the value of the real? I wasn't fully aware of this before the rise of LLMs, but it turns out that I am deeply committed to things being real. A simulation of X, for almost any putatively valuable X, has less moral worth, in my view, than the real thing. Why is that? Why will a Language Model Agent never be a real friend? Why do I want to stand before Hopper's Nighthawks myself, instead of seeing an infinite number of aesthetically equally-pleasing products of generative AI systems? I have some initial thoughts; but as AI systems become ever better at simulating everything that we care about, a fully-worked out theory of the value of the real, the authentic, will become morally and practically essential.

## Attention Guardians[36]

The pathologies of the digital public sphere derive in part from two problems. First, we unavoidably rely on AI to help us navigate the functionally infinite amount of online content. Second, existing systems for allocating online attention support the centralised, extractive power of a few big tech companies. Language Model Agents, functioning as Attention Guardians, could change this.

Our online attention is presently allocated using ML systems for recommendation and information retrieval that have three key features: they depend on vast amounts of behavioural data; they infer our preferences from our revealed behaviour; and they are controlled by private companies with little incentive to act in our interests. Deep reinforcement learning-based recommender systems, for example, are a fundamentally centralising and surveillant technology. Behavioural data must be gathered and centralised to be used to make inferences about relevance and irrelevance. Because this data is so valuable, and collecting it is costly, those who do so are not minded to share it—and because it is so potent, there are good data protection-based reasons not to do so.[37] As a result, only the major platforms are in a position to make effective retrieval and recommendation tools; their interests and ours are not aligned, leading to the practice of optimising for engagement, so as to maximise advertiser returns despite the individual and societal costs. And even if they aspired to actually advance our interests, RL permits only inferring revealed preferences—the preferences that we act on, not the preferences we wish we had. While the pathologies of online communication are obviously not all due to the affordances of recommender systems, this is an unfortunate mix.[38]

[34] Van Dijck, 2013.
[35] Chalmers, 2022.
[36] I discuss Attention Guardians in greater depth in Lazar, Forthcoming. For an initial proof of concept for a related idea, see Friedman et al., 2023.

Language Model Agents would enable Attention Guardians that differ in each respect. They would not depend on vast amounts of live behavioural data to function. They can (functionally) understand and operationalise your actual, not your revealed preferences. And they do not need to be controlled by the major platforms.

Obviously LLMs must be trained on tremendous amounts of data, but once trained they are highly adept at making inferences without ongoing surveillance. Imagine that data is blood. Existing deep RL-based recommender systems are like vampires, which must feed on the blood of the living to survive. Language Model Agents are more like combustion engines, relying on the oil produced by 'fossilised' data. Existing RL recommenders need centralised surveillance in order to model the content of posts online, to predict your preferences (by comparing your behaviour with others'), and so to map the one to the other. Language Model Agents could understand content simply by understanding content. And they can make inferences about what you would benefit from seeing using their reasoning ability and their model of your preferences, without relying on knowing what everyone else is up to.

This point is crucial: because of their facility with moral and related concepts, Language Model Agents could build a model of your preferences and values by directly talking about them with you, transparently responding to your actual concerns instead of just inferring what you like from what you do. This means that instead of bypassing your agency, they can scaffold it, helping you to honour your second-order preferences (about what you want to want), and learning from natural language explanations—even oblique ones—about why you don't want to see some particular post. And beyond just pandering to your preferences, Attention Guardians could be designed to be modestly paternalistic as well—in a transparent way.[39]

And because these Attention Guardians would not need behavioural data to function, and the infrastructure they depend on need not be centrally controlled by the major digital platforms, they could be designed to genuinely operate in your interests, and guard your attention instead of exploiting it.[40] While the major platforms would undoubtedly restrict Language Model Agents from browsing their sites on your behalf, they could transform the experience of using open protocol-based social media sites, like Mastodon, providing recommendation and filtering without surveillance and engagement-optimisation.

[37] Keller, 2021.
[38] On the relative role of algorithms vs societal factors, see e.g. Bail, 2021; Stray et al., 2022.
[39] See, for example, the proposal in Bernstein et al., 2023.
[40] Existing approaches to 'middleware' fall at this hurdle: they rely on behavioural data and so cannot be sufficiently independent from the platforms Keller, 2021. Note that I think Apple is a prime candidate for developing an Attention Guardian that is independent from the major platforms—it would be a similar move to their attempts to protect user privacy.

### Universal Intermediaries

Lastly, LLMs might enable us to design Universal Intermediaries, Language Model Agents sitting between us and our digital technologies, enabling us to simply voice an intention and see it effectively actualised by those systems. Everyone could have a digital butler, research assistant, personal assistant, and so on. The hierophantic coder class could be toppled, as everyone could conjure any program into existence with only natural language instructions.

At present, universal intermediaries are disbarred by LLMs' vulnerability to being hijacked by prompt injection. Because they do not clearly distinguish between commands and data, the data in their context window can be poisoned with commands directing them to behave in ways unintended by the person using them.[41] This is a deep problem—the more capabilities we delegate to Language Model Agents, the more damage they could do if compromised. Imagine an assistant that triages your email—if hijacked, it could forward all your private mail to a third party; but if we require user authorisation before the Agent can act, then we lose much of the benefit of automation.

But suppose these security hurdles can be overcome. Should we welcome universal intermediaries? I have written elsewhere that algorithmic intermediaries govern those who use them—they constitute the social relations that they mediate, making some things possible and others impossible, some things easy and others hard, in the service of implementing and enforcing norms.[42] Universal Intermediaries would be the apotheosis of this form, and would potentially grant extraordinary power to the entities that shape those intermediaries' behaviours, and so govern their users. This would definitely be a worry!

Conversely, if research on LLMs continues to make significant progress, so that highly capable Language Model Agents can be run and operated locally, fully within the control of their users, these Universal Intermediaries could enable us to autonomously govern our own interactions with digital technologies in ways that the centralising affordances of existing digital technologies render impossible. Of course, self-governance alone is not enough (we must also coordinate). But excising the currently ineliminable role of private companies would be significant moral progress.

## Conclusion

Existing Generative AI Systems are already causing real harms in the ways highlighted by the critics above. And future Language Model Agents—perhaps not the next generation, but before too long—may be dangerous enough to warrant at least some of the fears of looming AI catastrophe. But between these two extremes, the novel capabilities of the most advanced AI systems will enable a genre of Language Model Agents that is either literally unprecedented, or else has only been achieved in a piecemeal, inadequate way before. These new kinds of agents bring new urgency to previously neglected philosophical questions. Their societal impacts may be unambiguously bad, or there may be some good mixed in—in many respects, it is too early to say for sure, not only because we are uncertain about the nature of those effects, but because we lack adequate moral and political theories with which to evaluate them. It is now commonplace to talk about the design and regulation of 'frontier' AI models. If we're going to do either wisely, and build Language Model Agents that we can trust (or else decide to abandon them entirely), then we also need some frontier AI ethics.

[41] Greshake et al., 2023.
[42] Lazar, Forthcoming.